\documentclass{easychair}
\usepackage{doc}
\usepackage{indentfirst}
\usepackage{multirow}
\usepackage{algorithm}
\usepackage{algpseudocode}

\usepackage{cleveref}

\providecommand{\keywords}[1]
{
  \small	
  \textbf{\textit{Keywords---}} #1
}
\title{Robustness and Accuracy in Pipelined Bi-Conjugate Gradient Stabilized Method: A Comparative Study}
\author{
    Mykhailo Havdiak\inst{1}
\and 
Jos\'e I. Aliaga\inst{2}
\and
Roman Iakymchuk\inst{3} 
}

\institute{
Ivan Franko National University of Lviv, Ukraine\\ 
   \email{mykhailo.havdiak@lnu.edu.ua}
\and
   Universitat Jaime I, Spain\\
   \email{aliaga@icc.uji.es}
\and
  Umeå University and Uppsala University, Sweden\\ 
  \email{riakymch@cs.umu.se} 
 }

\authorrunning{}

\titlerunning{Robustness and Accuracy in Pipelined Biconjugate Gradient Stabilized Method: A Comparative Study}

\begin{document}

\maketitle

\begin{abstract}

\par In this article, we propose an accuracy-assuring technique for finding a solution for unsymmetric linear systems. Such problems are related to different areas such as image processing, computer vision, and computational fluid dynamics. Parallel implementation of Krylov subspace methods speeds up finding approximate solutions for linear systems. In this context, the refined approach in pipelined BiCGStab enhances scalability on distributed memory machines, yielding to substantial speed improvements compared to the standard  BiCGStab method. However, it’s worth noting that the pipelined BiCGStab algorithm sacrifices some accuracy, which is stabilized with the residual replacement technique. This paper aims to address this issue by employing the ExBLAS-based reproducible approach. We validate the idea on a set of matrices from the SuiteSparse Matrix Collection.

\end{abstract}


\setcounter{tocdepth}{2}
\section{Introduction}
\label{sect:introduction}

Krylov subspace methods form a powerful class of iterative techniques for solving large linear systems arising in diverse scientific and engineering applications. These methods are particularly well-suited for problems where the coefficient matrix is sparse and both symmetric or non-symmetric. Such methods are applicable and often used in image denoising, data compression, inverse problems, and other areas. The Conjugate Gradient (CG) method, introduced in ~\cite{Hestenes}, is one of the earliest members of this well-known class of iterative solvers. However, CG is limited to solving symmetric and positive definite (SPD) systems. In contrast, the Bi-Conjugate Gradient~\cite{Fletcher} (BiCG) method extends its applicability to more general classes of non-symmetric and indefinite linear systems. Additionally, the Conjugate Gradient Squared~\cite{Sonneveld} (CGS) method provides an alternative approach. 
Preconditioning is usually incorporated in real implementations of these methods in order to accelerate the convergence of the methods and improve their numerical features.

These classical Krylov subspace methods have been actively discussed and optimized. For instance, optimizations have been explored for a specific class of hepta-diagonal sparse matrices on GPUs~\cite{mod}, as well as the implementation of the pipelined Bi-Conjugate Gradient Stabilized method (p-BiCGStab)~\cite{COOLS20171} to overlap (hide) communication and computation. The pipelined methods, in particular, introduced more operations compared to the original ones and with that impacted the convergence as the computer residual deviated from the true one. As a remedy, the residual replacement technique was proposed~\cite{Carson:EECS-2012-197,COOLS20171} to numerically stabilize convergence with a strong emphasis on mathematical aspects. 

The purpose of this initial study is to explore the possibility of avoiding residual replacement in pipelined Krylov-type methods~\cite{COOLS20171} with the help of accurate and reproducible computations via the ExBLAS approach~\cite{Iakymchuk15ExBLAS,iakymchuk2023general}. As a test case, we use the pipelined BiCGStab method.

\section{Reproducibility of BiCGStab and pipelined BiCGStab}
\label{sect:bicgstab}
The BiConjugate Gradient Stabilized (BiCGStab) method~\cite{Vorst} was introduced as a smoother converging version of both BiCG and CGS methods. BiCGStab was developed to solve non-symmetric linear systems while avoiding the often irregular convergence patterns of the CGS method. In BiCGStab, minimizing a residual vector promotes smoother convergence. However, when the Generalized Minimal Residual method (GMRES)~\cite{gmres} stagnates, 
preventing the expansion of the Krylov subspace, BiCGStab may fail to proceed effectively.

\hspace*{-10mm}
\begin{minipage}{0.42\textwidth}
\vspace*{-33.6mm}
\begin{algorithm}[H]
\caption{BiCGStab~\cite{Vorst}}\label{alg:BiCGStab}
\begin{algorithmic}
\Function{BiCGStab}{A, b, $x_0$}
\State $r_0 := b - Ax_0$
\State $p_0 := r_0$

\For{i = 0, 1, 2, ...}{}{}
\State $s_i := Ap_i$
\State $a_i := (r_0, r_i) / (r_0, s_i)$
\State $q_i := r_i - a_i s_i$
\State $y_i := Aq_i$
\State $w_i := (q_i, y_i) / (y_i, y_i)$
\State $x_{i+1} := x_i + a_i p_i + w_i q_i$
\State $r_{i+1} := q_i - w_i y_i$
\State $\beta_i := (a_i/w_i)(r_0, r_{i+1})/(r_0, r_i)$
\State $p_{i+1} := r_{i+1} + \beta_i(p_i - w_i s_i)$
\EndFor

\EndFunction
\end{algorithmic}
\end{algorithm}
\end{minipage}
\begin{minipage}{0.57\textwidth}
\begin{algorithm}[H]
\caption{Pipelined BiCGStab (p-BiCGStab)~\cite{COOLS20171}}\label{alg:p-BiCGStab}
\begin{algorithmic}
\Function{p-BiCGStab}{A, b, $x_0$}
\State $r_0 := b - Ax_0$
\State $w_0 := A - r_0$
\State $t_0 := Aw_0$
\State $a_0 := (r_0, r_0) / (r_0, w_0)$
\State $\beta_{-1} := 0$
\For{i = 0, 1, 2, ...}{}{}
\State $p_i := r_i + \beta_{i-1}(p_{i-1} - w_{i-1} s_{i-1})$
\State $s_i := w_i + \beta_{i-1}(s_{i-1} - w_{i-1} z_{i-1})$
\State $z_i := t_i + \beta_{i-1}(z_{i-1} - w_{i-1} v_{i-1})$
\State $q_i := r_i - a_i s_i$
\State $y_i := w_i - a_i z_i$
\State $v_i := Az_i$
\State $w_i := (q_i, y_i)/(y_i, y_i)$
\State $x_{i+1} := x_i + a_i p_i + w_i q_i$
\State $r_{i+1} := q_i - w_i y_i$
\State $w_{i+1} := y_i - w_i (t_i - a_i v_i)$
\State $t_{i+1} := Aw_{i+1}$
\State $\beta_i := (a_i/w_i)(r_0, r_{i+1})/(r_0, r_i)$
\State $a_{i+1} := (r_0, r_{i+1})/((r_0, w_{i+1}) + \beta_i(r_0, s_i) - \beta_i w_i (r_0, z_i))$
\EndFor
\EndFunction
\end{algorithmic}
\end{algorithm}
\end{minipage}
\vspace*{4mm}

In the light of the conventional BiCGStab algorithm, see~\Cref{alg:BiCGStab}, introduced by H.A. Van der Vorst, S. Cools and Wim Vanroose proposed an optimization known as pipelined BiCGStab (p-BiCGStab)~\cite{COOLS20171}, see~\Cref{alg:p-BiCGStab}. This optimization entails two primary phases within the pipelining framework. Firstly, in what is termed the `communication-avoiding' phase, the standard Krylov algorithm undergoes a transformation into a mathematically equivalent form with the reduced global synchronization points. This reduction is accomplished by merging the global reduction phases of various dot products scattered throughout the algorithm into a single global communication phase. Subsequently, in the `communication-hiding' phase, the algorithm is further refined to overlap the remaining global reduction phases with the sparse matrix-vector product and application of the preconditioner. This strategic restructuring effectively mitigates the typical communication bottleneck by concealing communication time behind productive computational tasks. Although, the methods are mathematically equivalent they may lead to different numerical results and convergence patterns due to the non-associativity of floating-point operations.


The BiCGStab (\Cref{alg:BiCGStab}) and p-BiCGStab (\Cref{alg:p-BiCGStab}) methods will serve as the primary methods utilized throughout this article, although we mainly focus on the p-BiCGStab. Due to the non-associativity of finite precision floating-point operations, the mathematical equivalent of these two methods can show large divergence while implemented in parallel environments especially for tolerance below $10^{-6}$. To stabilize this deviation in p-BiCGStab, the residual replacement technique was proposed~\cite{Carson:EECS-2012-197}. This technique resets the residuals $r_i$, along with the auxiliary variables $w_i$, $s_i$, and $z_i$, to their original values every $k$ iterations. In the preconditioned version, this process also updates $\bar{r_i} = M^{-1}r_i$ and $\bar{s_i} = M^{-1}s_i$, where $M$ is the preconditioning operator.

In~\cite{iakymchuk2023general}, we proposed to ensure the reproducibility and accuracy of the pure MPI implementation of the preconditioned BiCGStab method via the ExBLAS approach. ExBLAS combines together long accumulator and floating-point expansions into algorithmic solutions as well as efficiently tunes and implements them on various architectures. ExBLAS aims to provide new algorithms and implementations for fundamental linear algebra operations (like those included in the BLAS library), that deliver reproducible and accurate results with small or without losses to their performance on modern parallel architectures such as desktop and server processors, Intel Xeon Phi co-processors, and GPU accelerators. We construct our approach in such a way that it is independent of data partitioning, order of computations, thread scheduling, or reduction tree schemes. Instead of using the residual replacement technique, we propose to exhibit the benefits of the ExBLAS approach in the pipelined BiCGStab method.

\section{Experimental Results}
\label{sect:exp}

This section presents a series of numerical experiments to evaluate the convergence, performance, and accuracy of the BiCGStab methods, including the reproducible one with ExBLAS. The results include comparisons between BiCGStab, pipelined BiCGStab (p-BiCGStab), p-BiCGStab with ExBLAS, and p-BiCGStab with the residual replacement technique across various matrices sourced from the Suite Sparse Matrix Collection. In the experiments, IEEE754 double-precision arithmetic was utilized, and we run on the dual 14-core Intel Xeon Gold 6132 (Skylake) @2.60GHz at HPC2N. The nodes are interconnected via EDR Infiniband.


The SuiteSparse Matrix Collection~\cite{suitesparse} is a comprehensive repository of sparse matrices widely used for benchmarking and testing numerical algorithms in the field of computational mathematics. It allows for robust and repeatable experiments, as performance results with artificially generated matrices can be misleading. Hence, repeatable experiments are crucial for ensuring the reliability of algorithm evaluations. The collection encompasses a diverse range of matrices representing real-world problems from various disciplines.


\Cref{tab:1} presents the comparative performance of four iterative BiCGStab methods, namely BiCGStab, pipelined BiCGStab (p-BiCGStab), p-BiCGStab with ExBLAS, and pipelined BiCGStab with residual replacement (p-BiCGStabRR), across a selection of sparse matrices from the SuiteSparse Matrix Collection. Each cell in the table represents the number of iterations required to achieve convergence for a specific method and a given matrix, with convergence thresholds set at $10^{-6}$ and  $10^{-9}$. The results demonstrate varying convergence behavior among the methods across different matrices, providing insights into their respective efficiency in solving sparse linear systems. 

\begin{table}[ht]
\centering
\begin{tabular}{|l|cc|cc|cc|cc|}
\hline
 \multirow{2}{4em}{Problem}    &  \multicolumn{2}{|c|}{BiCGStab}   & \multicolumn{2}{|c|}{p-BiCGStab} & \multicolumn{2}{|c|}{p-BiCGStabExBLAS} &  \multicolumn{2}{c|}{p-BiCGStabRR} \\
  &   $1e-6$ &   $1e-9$ &   $1e-6$ &   $1e-9$ & $1e-6$   &   $1e-9$ &   $1e-6$ & $1e-9$ \\
\hline \rule{0pt}{2ex}
 1138\_bus &   30 &   151  &   \textbf{27}  &  130 &   35  &  \textbf{108} &  \textbf{27}  & 130  \\ \rule{0pt}{2ex}
 add32    &    \textbf{38} &   74   &   \textbf{38}  &   \textbf{68} &   38  &   \textbf{69} &  38  & \textbf{68}   \\ \rule{0pt}{2ex}
 bcsstk13 &   545 &   -    &  520  & 2258 &  350  & 3273 & \textbf{195}  & \textbf{403}  \\ \rule{0pt}{2ex}
 bcsstk14 &   149 &   461  &   44  &  459 &   \textbf{43}  &  \textbf{433} &  44  & -    \\ \rule{0pt}{2ex}
 bcsstk18 &   405 &   2806 &  \textbf{261}  & 1284 &  366  & \textbf{1274} & 309  & -    \\ \rule{0pt}{2ex}
 bcsstk27 &   283 &   \textbf{958}  &  335  & 2107 &  \textbf{279}  & 1477 & 335  & 2107 \\ \rule{0pt}{2ex}
 bfwa782  &    99 &   647  &   74  &  \textbf{448} &   \textbf{54}  &  463 & 115  & 576  \\ \rule{0pt}{2ex}
 cdde6    &    36 &   122  &   \textbf{34}  &  121 &   \textbf{34}  &  \textbf{115} &  \textbf{34}  & 388  \\ \rule{0pt}{2ex}
 msc01050 &    \textbf{29} &   61   &   30  &   \textbf{47} &   \textbf{28}  &   60 &  30  & \textbf{47}   \\ \rule{0pt}{2ex}
 msc04515 &   123 &   \textbf{257}  &   \textbf{96}  &  275 &   \textbf{98}  &  308 & 263  & 334  \\ \rule{0pt}{2ex}
 orsreg\_1 &   \textbf{21} &   161  &   22  &  168 &   \textbf{20}  &  \textbf{106} &  22  & 371  \\ \rule{0pt}{2ex}
 pde2961  &   \textbf{100} &   \textbf{166}  &  111  &  278 &  134  &  287 & 170  & 683  \\ \rule{0pt}{2ex}
 plat1919 &    \textbf{79} &   \textbf{132}  &   99  &  185 &   84  &  179 &  87  & 250  \\ \rule{0pt}{2ex}
 rdb3200l &    31 &   193  &   31  &  223 &   31  &  \textbf{171} &  31  & 567  \\ \rule{0pt}{2ex}
 saylr4   &    28 &   73   &   28  &   74 &   28  &   \textbf{69} &  28  & 74   \\ \rule{0pt}{2ex}
 sherman3 &    34 &   501  &   33  &  314 &   \textbf{26}  &  400 &  33  & \textbf{271}  \\ \rule{0pt}{2ex}
 utm5940  &    99 &   592  &   97  &  \textbf{420} &   \textbf{18}  &  603 &  20  & \textbf{419}  \\ 
\hline
\end{tabular}
\caption{Number of iterations for the BiCGStab-like methods on a set of the SuiteSparse matrices without precondition, where fewer iterations indicate better performance. The initial estimate is set to $x_0=0$. The best-performing method is highlighted in bold.}
\label{tab:1}
\vspace*{-4mm}
\end{table}

When summarizing the findings, several notable observations come to light. Firstly, for $\varepsilon = 10^{-6}$, all methods demonstrate comparable performance in numerous cases. However, when considering a higher tolerance, $\varepsilon = 10^{-9}$, the p-BiCGStabExBLAS method consistently outperforms p-BiCGStab across the majority of matrices, owing to its enhanced accuracy. Moreover, the p-BiCGStab with residual replacement strategy (p-BiCGStabRR) method generally exhibits superior convergence rates compared to the p-BiCGStab method. Yet, increasing the tolerance also reveals instances where the classic BiCGStab method proves more efficient in terms of iterations, although it may encounter convergence issues for certain matrices. Notably, for the specific {\tt bcsstk13} problem, p-BiCGStabRR demonstrates the most favorable convergence characteristics for both $tol = 10^{-6}$, $tol = 10^{-9}$. Despite these advantages, there are scenarios where p-BiCGStabRR fails to converge, namely {\tt bcsstk14} and {\tt bcsstk18}. The accuracy of the p-BiCGStabRR is highly contingent to the specific problem context and parameter choices, leading to variability in its effectiveness. This method may exhibit convergence speed under certain conditions while performing poorly under others, highlighting the sensitivity of its outcomes to these factors. The method requires multiple runs to determine the optimal step and the best place for applying residual replacement. A less optimal parameter choice can result in more iterations compared to the pipelined-BiCGStab method.

\begin{figure}[ht]
\centering
\includegraphics[scale=0.35]{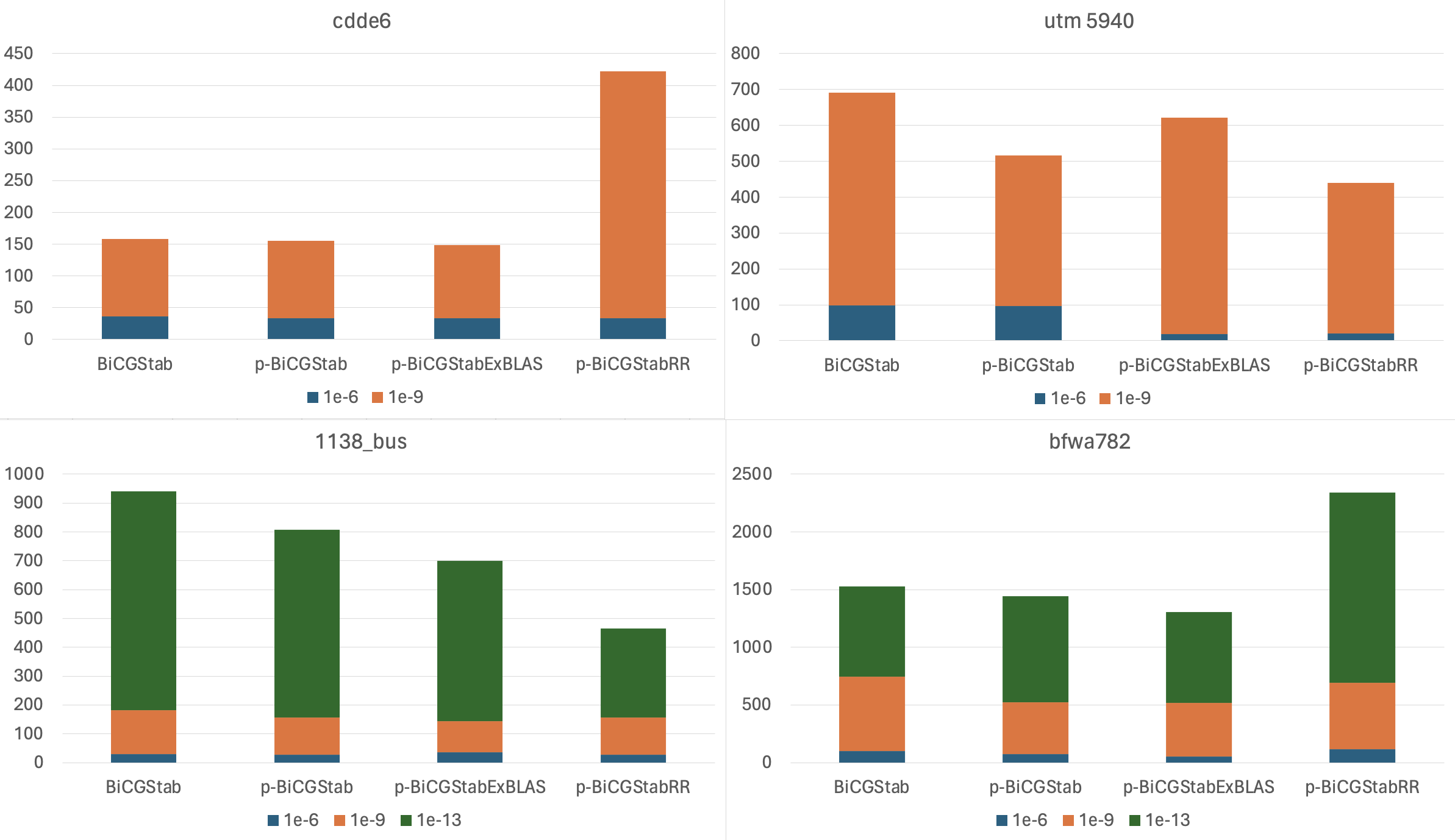}
\vspace*{-5mm}
\caption{Number of iterations required by various BiCGStab-like methods to achieve a specified tolerance ($10^{-6}, 10^{-9}, 10^{-13}$). p-BiCGStabRR stands for the pipelined version of the BiCGStab method with residual replacement; P-BiCGStabExBLAS refers to the method with ExBLAS.}. 
\label{fig1}
\vspace*{-6mm}
\end{figure}  

In~\Cref{fig1}, the {\tt utm5940} case highlights an interesting trend: the ExBLAS version performed the best for epsilon $10^{-6}$, yet with an increase to $10^{-9}$, it required slightly more iterations compared to other methods. Overall, p-BiCGStabExBLAS demonstrates good constant performance in terms of iterations. When examining p-BiCGStabRR, it's evident that for certain examples, it exhibits the lowest iteration count. However, there are instances for higher tolerance the method 
results in significantly higher iteration counts compared to other methods.

\begin{figure} [!ht]
\centering
\vspace*{-4mm}
\hspace*{-4mm}\includegraphics[scale=0.18]{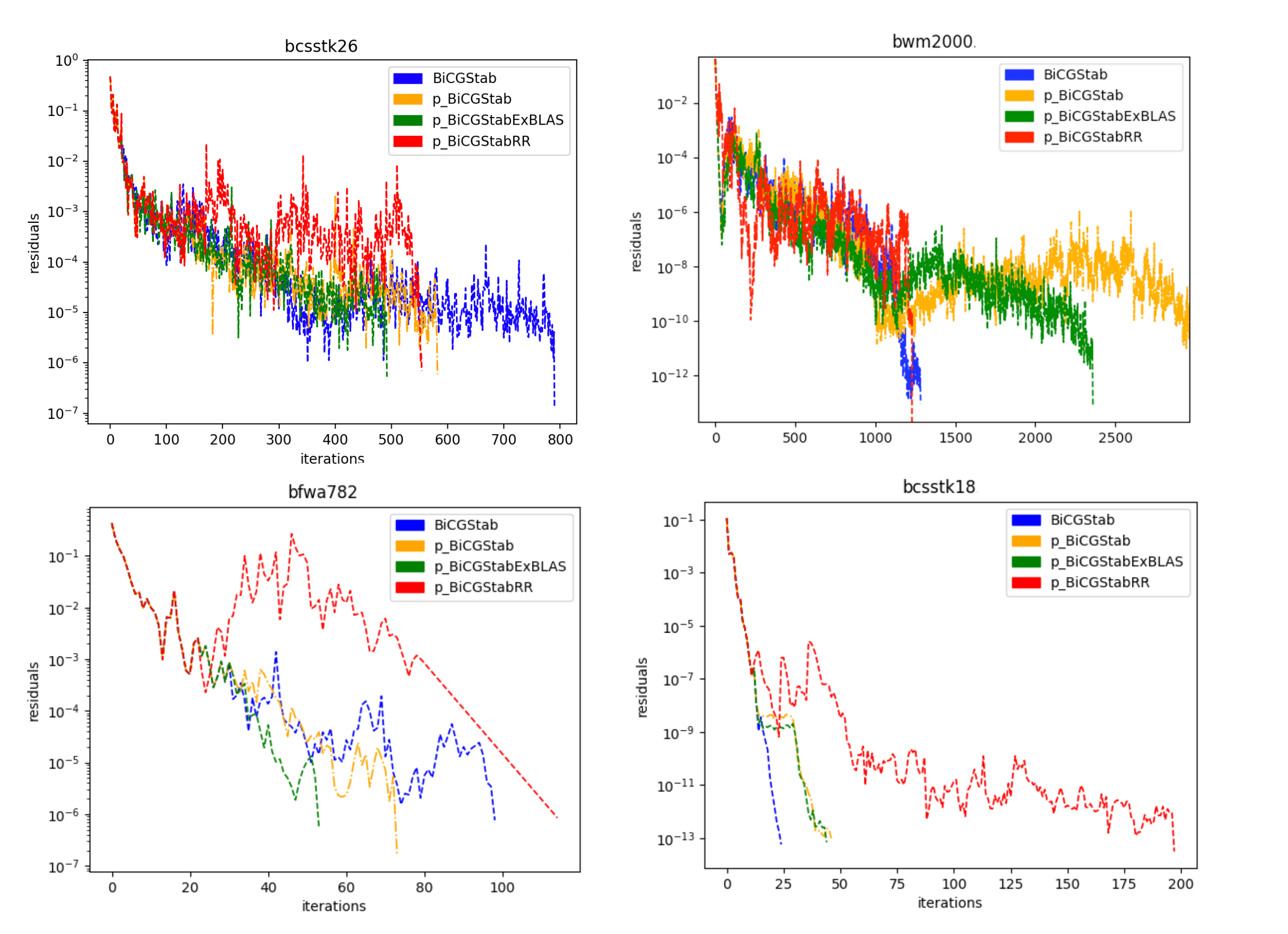}
\vspace*{-5mm}
\caption{Residual history of the four BiCGStab-like methods: BiCGStab, pipelined BiCGStab (p-BiCGStab), pipelined BiCGStab with ExBLAS (p-BiCGStabExBLAS), pipelined BiCGStab with residual replacement (p-BiCGStabRR); $tol = 10^{-13}$.}
\label{fig2}
\vspace*{-6mm}
\end{figure}  

\Cref{fig2} provides the convergence history of the four BiCGStab-like methods. We can observe the performance characteristics of the methods on a particular problem instance, where the tolerance is set to $10^{-13}$. The pipelined BiCGStab method with ExBLAS consistently outperforms the regular pipelined variant in terms of iterations across a wide range of scenarios. p-BiCGStabExBLAS exhibits a reduced occurrence of spikes compared to p-BiCGStabRR, suggesting a smoother and more stable performance profile. This difference highlights the potential of ExBLAS to offer improved reliability and predictability in computational outcomes, namely results and iterations.

\begin{table}[ht]
\centering
\begin{tabular}{|l|ccc|ccc|ccc|}
\hline
\multirow{2}{4em}{$\varepsilon = 10^{-6}$} & \multicolumn{3}{|c|}{BiCGStab} & \multicolumn{3}{|c|}{p-BiCGStab} & \multicolumn{3}{|c|}{p-BiCGStabExBLAS} \\ \rule{0pt}{2ex}
 &  $n=1$ & $n=8$ & $n=16$ & $n=1$ & $n=8$ & $n=16$&$n=1$ & $n=8$ & $n=16$\\ 
\hline  \rule{0pt}{2ex}
bcsstk26 & 791 & 235 & 599 & 583 & 528 & 683 & 493 & 493 & 493 \\ \rule{0pt}{2ex}
bwm2000 & 37 & 37 & 37 & 37 & 37 & 37 & 37 & 37 & 37 \\ \rule{0pt}{2ex}
bfwa782 & 99 & 79 & 85 & 74 & 104 & 59 & 54 & 54 & 54 \\ \rule{0pt}{2ex}
bcsstk18 & 405 &  414 & 363 & 261 & 305 & 356 & 366 & 366 & 366 \\
\hline
\end{tabular}
\caption{Number of iterations required for BiCGStab, p-BiCGStab, and p-BiCGStabExBLAS with varying numbers of processes ($n$) for $\varepsilon = 10^{-6}$.}
\label{tab:2}
\vspace*{-5mm}
\end{table}
\begin{table}[ht]
\centering
\begin{tabular}{|l|ccc|ccc|ccc|}
\hline
\multirow{2}{4em}{Matrix} & \multicolumn{3}{|c|}{BiCGStab} & \multicolumn{3}{|c|}{p-BiCGStab} & \multicolumn{3}{|c|}{p-BiCGStabExBLAS} \\ \rule{0pt}{2ex}
 &  $n=1$ & $n=8$ & $n=16$ & $n=1$ & $n=8$ & $n=16$&$n=1$ & $n=8$ & $n=16$\\ 
\hline  \rule{0pt}{2ex}
{\tt bwm2000} & 1268 & 1267 & 1120 & 663 & 1131 & 1024 & 1232 & 1232 & 1232 \\ \rule{0pt}{2ex}
{\tt bfwa782} &  647 & 528 & 531 & 448 & 476 & 498 & 464 & 463 & 463 \\ \rule{0pt}{2ex}
{\tt bcsstk18} & 2806 & 2819 & 2286 & 1284 & 1811 & 2318 & 1274 & 1274 & 1274 \\
\hline
\end{tabular}
\caption{Number of iterations required for BiCGStab, p-BiCGStab, and p-BiCGStabExBLAS by varying the numbers of processes ($n$) for $tol = 10^{-9}$.}
\label{tab:3}
\vspace*{-3mm}
\end{table}

Following this, we evaluate the four considered methods using an increased number of processes. Subsequently, we present \Cref{tab:2,tab:3}  illustrating the outcomes achieved by the aforementioned methods across different process counts.
A notable observation from both~\Cref{tab:2} and \Cref{tab:3} is the consistency in the number of iterations required for the ExBLAS implementation across different numbers of processes. Thus, increasing the number of processes does not lead to a faster solution for pipelined BiCGStab method.
The pipelined BiCGStabRR as indicated in~\Cref{tab:1} demonstrates its best outcome of 195 iterations for the {\tt bcsstk13} matrix. 

\begin{table}[!ht]
\centering
\begin{tabular}{|l|cc|cc|cc|}
\hline  \rule{0pt}{2ex}
\multirow{2}{4em}{Method}  & \multicolumn{2}{|c|}{$n = 1$}  & \multicolumn{2}{|c|}{$n = 8$} & \multicolumn{2}{|c|}{$n = 16$}  \\

& iter & time & iter & time & iter & time \\
\hline  \rule{0pt}{3ex}
BiCGStab & 545  & $2.034 \times 10^{-1}$ & 520 & $4.8236 \times 10^{-2}$ & 544 & $3.94\times 10^{-2}$\\ \rule{0pt}{2ex}
p-BiCGStab & 520 & $2.0495 \times 10^{-1}$ & 482 & $4.457 \times 10^{-2}$ & 394 & $3.106 \times 10^{-2}$ \\ \rule{0pt}{2ex}
p-BiCGStabE & 350 & $5.321 \times 10^{-1}$& 350 & $8.976 \times 10^{-2}$ & 350 & $5.819 \times 10^{-2}$ \\ \rule{0pt}{2ex}
p-BiCGStabRR & 195 & $6.198 \times 10^{-2}$ &  \multicolumn{2}{|c|}{-}  & \multicolumn{2}{|c|}{-} \\
\hline
\end{tabular}
\caption{Number of iterations and time required for the BiCGStab-like methods for the {\tt bcsstk13} matrix. Tolerance is set to $tol = 10^{-6}$.}
\label{tab:4}
\end{table}

\Cref{tab:4} illustrates the iteration counts for the BiCGStab, p-BiCGStab, and p-BiCGStabExBLAS methods across different numbers of processes.
The method with residual replacement technique finds an approximation to the solution only on a single process. Additionally, the table presents the execution times for each scenario. p-BiCGStabExBLAS requires more time, attributed to its higher accuracy. Nonetheless, the overhead associated with p-BiCGStabExBLAS diminishes as the number of processes increases, dropping from 2.6x on a single process to 1.87x on 16 processes.

\begin{figure} [!ht]
\centering
\vspace*{-3mm}
\hspace*{-4mm}\includegraphics[scale=0.30]{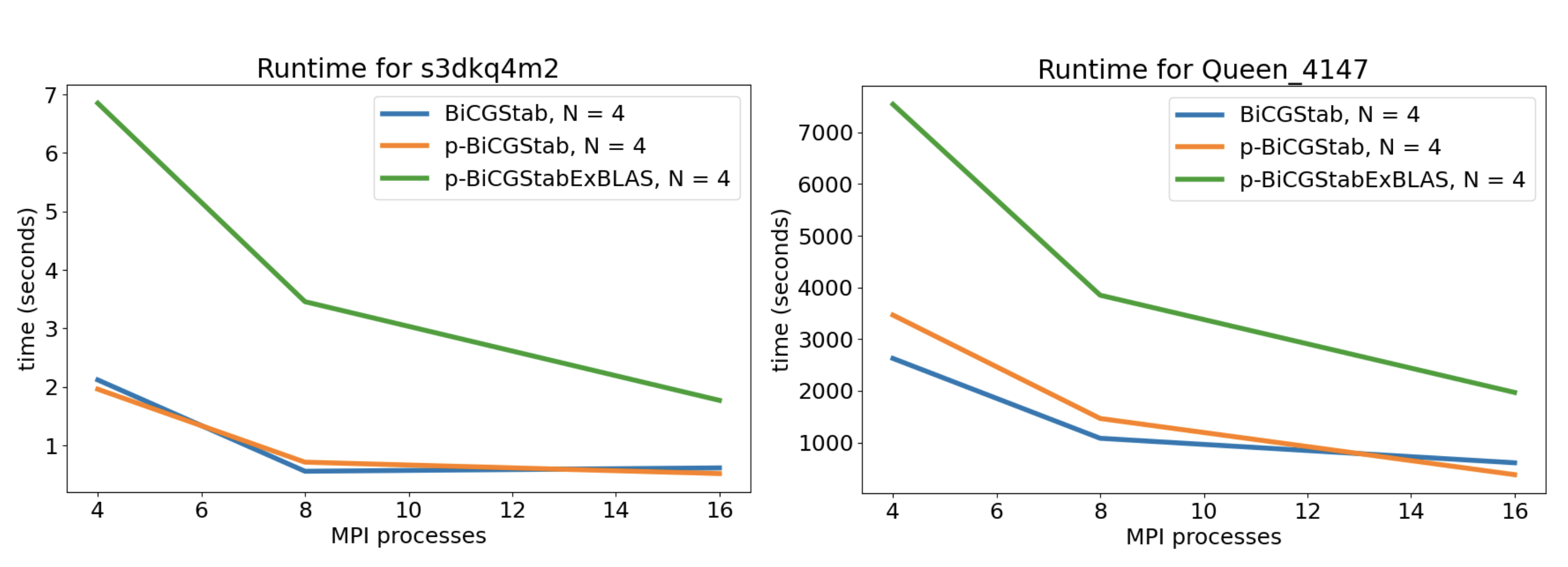}
\vspace*{-7mm}
\caption{
Runtime comparison of BiCGStab-like methods on two SuiteSparse matrices, namely Queen\_4147 with $316,548,962$~nnz and s3dkq4m2 with $4,427,725$~nnz, using a various number of MPI processes; $tol=10^{-6}$ for Queen\_4147 and $tol=10^{-9}$ for s3dkq4m2.}
\label{fig3}
\vspace*{-1mm}
\end{figure}
\Cref{fig3} illustrates the benefits of using pipelined methods within a parallel environment, emphasizing their efficiency and scalability. Certainly, the scale is small but the gain starts to be visible on 16 processes, four per each of four nodes; we used only few processes per node to highlight the benefit. In this test, we focus on two large matrices: {\tt s3dkq4m2} with $4,427,725$ non-zero elements and {\tt Queen\_4147} with $316,548,962$ non-zero elements. Larger-dimensional problems tend to demonstrate better strong scalability in parallel environments, using the existing potential of available resources especially on four and eight processes. 
Conversely, employing 16 processes for the {\tt s3dkq4m2} matrix with fewer non-zero elements did not yield significant improvements for BiCGStab and p-BiCGStab. Additionally, p-BiCGStabExBLAS demonstrates strong scalability for both problems due to more flops imposed by the ExBLAS approach. Overhead for s3dkq4m2 varied from 3.41x to 4.8x, and matrix Queen\_4147 from 1.96x to 5.2x. With the increase in the number of processes from 4 to 16, the execution time for p-BiCGStabExBLAS was reduced by more than 2x.

\section{Conclusion}
\label{sect:conclusion}

In this study, we investigated the robustness and accuracy of the pipelined Biconjugate Gradient Stabilized using the ExBLAS approach as not only an accurate and reproducible solution but also as an alternative to the residual replacement technique.
Our analysis focused on evaluating the convergence behavior of the method across a set of matrices from the SuiteSparse Matrix Collection. Through the numerical experiments, we demonstrated that the pipelined BiCGStab method with ExbLAS, consistently outperforms the conventional pipelined BiCGStab approach in terms of convergence rates and numerical reliability. Overall, this study emphasizes the importance of considering algorithmic refinements and numerical stability enhancements to achieve reliable and efficient solutions for challenging computational problems. The results underscore notable performance disparities among the assessed methods. Specifically, BiCGStab demonstrates better stability compared to p-BiCGStab, showcasing its reliability in solving linear systems. The residual replacement strategy is expected to address the stability of the pipelined method, bringing it closer to the robustness exhibited by BiCGStab. Although its performance is inferior to ExBLAS implementation. This suggests that the ExBLAS implementation capitalizes on the advantages of the pipelined method version while maintaining stability as in BiCGStab.

As a future work, 
we shall conduct theoretical study of the ExBLAS approach as a possible replacement for the residual replacement, which requires some empirical trials to get the right step. Furthermore, we plan to carry out an exhaustive study with more matrices from the SuiteSparse Matrix Collection as well as the real applications like the ones from the EU-funded EuroHPC JU Center of Excellence in Exascale CFD (CEEC)\footnote{This work was partially supported by CEEC under grant agreement No 10109339}, where the last author contributes.


\end{document}